\documentclass[conference, letterpaper, 10pt]{IEEEtran}
\usepackage[utf8]{inputenc}
\usepackage[pdftex,bookmarks=false]{hyperref}
\IEEEoverridecommandlockouts
\usepackage{amsmath,amssymb,amsfonts}
\usepackage{algorithmic}
\usepackage{graphicx}
\usepackage{textcomp}
\usepackage{xcolor}

\usepackage{booktabs}
\usepackage{stfloats}

\usepackage[ruled, vlined, linesnumbered, longend, resetcount]{algorithm2e}

\SetCommentSty{mycommfont}

\usepackage{amsmath}
\usepackage{mathtools}
\usepackage{tabularx}
\usepackage{amsfonts}

\usepackage{graphicx}
\usepackage{lipsum}
\usepackage{float}
\usepackage{afterpage}
\usepackage{adjustbox}

\usepackage[caption=false,font=normalsize,labelfont=sf,textfont=sf]{subfig}

\usepackage{orcidlink}
\usepackage{authblk}
\usepackage{placeins}

\usepackage[backend=biber, style=ieee]{biblatex}
\renewbibmacro{in:}{\ifentrytype{article}{}{\printtext{\bibstring{in}\intitlepunct}}}

\begin{document}

\title{Q-Learning-Based Time-Critical Data Aggregation Scheduling in IoT \\

\thanks{This work was partly supported by the Korea government (MSIT), IITP, Korea, under the ICT Creative Consilience program (IITP-2025-RS-2020-II201821, 20\%); AI Innovation Hub (RS-2021-II212068, 20\%), the Development of 6G Network Integrated Intelligence Plane Technologies (IITP-2025-RS-2024-00392332, 30\%); and the National Research Foundation of Korea (RS-2024-00343255, 30\%).}
}

\author[1]{Van-Vi Vo  \orcidlink{0000-0002-5745-4164}}
\author[2]{Tien-Dung Nguyen \orcidlink{0000-0003-0064-4044}}
\author[3,*]{Duc-Tai Le \orcidlink{0000-0002-5286-6629}}
\author[3,*]{Hyunseung Choo \orcidlink{0000-0002-6485-3155}}
\affil[1]{{\small Convergence Research Institute, Sungkyunkwan University, Suwon, South Korea}}
\affil[2]{{\small School of Electrical and Electronic Engineering, Hanoi University of Science and Technology, Hanoi, Vietnam}}
\affil[3]{{\small Dept. of Electrical and Computer Engineering, Sungkyunkwan University, Suwon, South Korea}}
\affil[*]{{\small Corresponding authors (\{ldtai, choo\}@skku.edu)}}


\maketitle

\begin{abstract}
Time-critical data aggregation in Internet of Things (IoT) networks demands efficient, collision-free scheduling to minimize latency for applications like smart cities and industrial automation. Traditional heuristic methods, with two-phase tree construction and scheduling, often suffer from high computational overhead and suboptimal delays due to their static nature. To address this, we propose a novel Q-learning framework that unifies aggregation tree construction and scheduling, modeling the process as a Markov Decision Process (MDP) with hashed states for scalability. By leveraging a reward function that promotes large, interference-free batch transmissions, our approach dynamically learns optimal scheduling policies. Simulations on static networks with up to 300 nodes demonstrate up to 10.87\% lower latency compared to a state-of-the-art heuristic algorithm, highlighting its robustness for delay-sensitive IoT applications. This framework enables timely insights in IoT environments, paving the way for scalable, low-latency data aggregation.

\end{abstract}


\section{Introduction}
The Internet of Things (IoT) ecosystem relies heavily on Wireless Sensor Networks (WSNs) to enable real-time data collection, monitoring, and decision-making across diverse applications such as smart cities, industrial automation, and environmental surveillance. WSNs consist of distributed sensor nodes that gather data and transmit it to a central sink, forming the backbone of IoT connectivity. However, WSNs face significant challenges in energy efficiency and network congestion, as sensor nodes are typically battery-powered with limited computational resources, making energy conservation critical to prolong network lifetime. Moreover, high data volumes from numerous nodes can cause network congestion, leading to delays and packet losses critical in time-sensitive scenarios~\cite{saadouni2025id, zaman2024review}. 

Data aggregation scheduling is pivotal in mitigating these issues by fusing redundant data at intermediate nodes, reducing transmission overhead and latency. In time-critical IoT applications, minimizing the aggregation delay, the number of timeslots required for all data to reach the sink, is essential to ensure timely insights and responses \cite{vo2024gnn, vo2024active}. Effective scheduling must also guarantee collision-free transmissions, adhering to interference models that prevent simultaneous transmissions by neighboring nodes. This process not only conserves energy but also enhances network performance, particularly in dense or asymmetric topologies.

Traditional heuristic algorithms, such as tree-based methods, often employ a two-phase approach: first constructing an aggregation tree and then scheduling transmissions along its paths. For instance, the break-and-join tree construction iteratively refines aggregation trees to minimize depth and delay \cite{nguyen2020break}, but while effective in static environments and sparse networks, these methods incur high computational overhead and may yield suboptimal delays due to decoupled optimization stages and their static nature, which limits performance in dynamic settings. Recent machine learning approaches, particularly reinforcement learning, have gained traction for adaptability; for example, Q-learning-based routing protocols integrate data aggregation awareness for energy-efficient routing \cite{yun2021q}. Recent works optimize information freshness in multi-hop IoT networks \cite{wu2025data} or employ blockchain with fuzzy similarity matrices to secure energy-efficient aggregation, focusing on clustering to minimize redundancy and energy dissipation \cite{ahmed2022energy}, yet these often prioritize energy, security, or freshness over strict latency constraints and rarely integrate tree formation with scheduling.

This paper proposes a Q-learning framework that unifies aggregation tree construction and scheduling for static IoT networks, modeling the process as a Markov Decision Process (MDP) with hashed states for scalability. Unlike prior works, our approach directly targets minimal aggregation delay, achieving up to 10.87\% lower latency compared to the break-and-join heuristic \cite{nguyen2020break}. The contributions of this work are threefold: 
\begin{itemize}
    \item A novel Q-learning framework unifying aggregation tree construction and scheduling, with a scalable hashed state representation.
    \item A greedy-augmented action selection mechanism with a batch-promoting reward function, ensuring efficient, collision-free batch transmissions.
    \item Extensive empirical evaluation demonstrating robust latency reduction, particularly in asymmetric topologies like corner sink configurations.
\end{itemize}

The remainder of the paper is structured as follows. Section \ref{network_stat} presents the network modeling and assumptions. Section \ref{proposed_appr} details the proposed Q-learning-based scheduling method. We evaluate the performance in Section \ref{exp_results}, conclude, and plan our work in Section \ref{conclusion}.

\section{Network Model and Problem Statement} \label{network_stat}
\subsection{Network modeling and Assumptions}
The IoT network is modeled as an undirected unit disk graph $G = (V, E)$, where $V = \{v_0, v_1, \dots, v_N\}$ includes the sink (node $v_0$) and $N$ sensor nodes randomly deployed in a $100 \times 100$ plane. An edge $(u, v) \in E$ exists if the Euclidean distance between nodes $u$ and $v$ is at most the communication range $R$, reflecting wireless connectivity constraints. All nodes operate with identical transmission power, and the sink aggregates data from all sensors for processing. This graph-based model captures network connectivity and interference patterns, enabling efficient scheduling of time-critical data aggregation.

To focus on latency minimization, we adopt several key assumptions. The network topology remains static throughout the aggregation cycle, with no node mobility or failures, ensuring consistent connectivity. Each sensor node generates exactly one unit of data per cycle, which intermediate aggregators can perfectly fuse into a single output packet, regardless of the number of inputs, to minimize transmission overhead. Transmissions occur in discrete time slots, allowing multiple non-interfering sends per slot. We employ the protocol interference model, where a receiver can process data from only one sender per slot, so no neighboring nodes of the receiver transmit simultaneously to prevent collisions, ensuring reliable data aggregation.

\subsection{Problem Formulation}
The Minimum Latency Aggregation Scheduling (MLAS) problem seeks to determine a schedule $S = \{S_1, S_2, \dots, S_D\}$ that minimizes the total number of time slots $D$, where each $S_t$ is a set of non-conflicting (sender, aggregator) pairs. During each time slot $t$, senders in $S_t$ transmit their aggregated data to their designated aggregators, adhering to interference constraints, until all sensor data reaches the sink. The MLAS problem is NP-hard under the protocol interference model~\cite{chen2005minimum}, necessitating scalable and efficient solutions for practical IoT deployments. To address this, we formulate the scheduling task as a Markov Decision Process (MDP), where states represent subsets of nodes whose data has been aggregated to the sink (directly or via intermediate nodes), actions involve selecting initial senders to initiate batch transmissions, transitions update the aggregated node set based on a greedy expansion mechanism, and rewards are designed to maximize interference-free batch sizes to reduce $D$.

\subsection{Problem Statement}
The objective is to assign time slots to nodes to achieve the following:
\begin{itemize}
    \item Minimize the number of time slots $D$ required for all sensor nodes' data to be aggregated and delivered to the sink.
    \item Ensure interference-free transmissions, such that no node receives from multiple senders in the same slot, and no sender transmits if it interferes with a nearby receiver (i.e., a neighbor of another receiver in the same slot).
    \item Enforce dependency constraints, ensuring a node transmits only after receiving data from its children in the implicit aggregation tree formed by the schedule.
\end{itemize}

This formulation ensures that the scheduling process is collision-free and latency-optimal, addressing the constraints of time-critical IoT networks while leveraging the graph-based model to capture spatial and interference relationships effectively.

\section{Proposed Approach} \label{proposed_appr}
\begin{figure}[htbp]
    \centering
    \includegraphics[width=0.9\columnwidth]{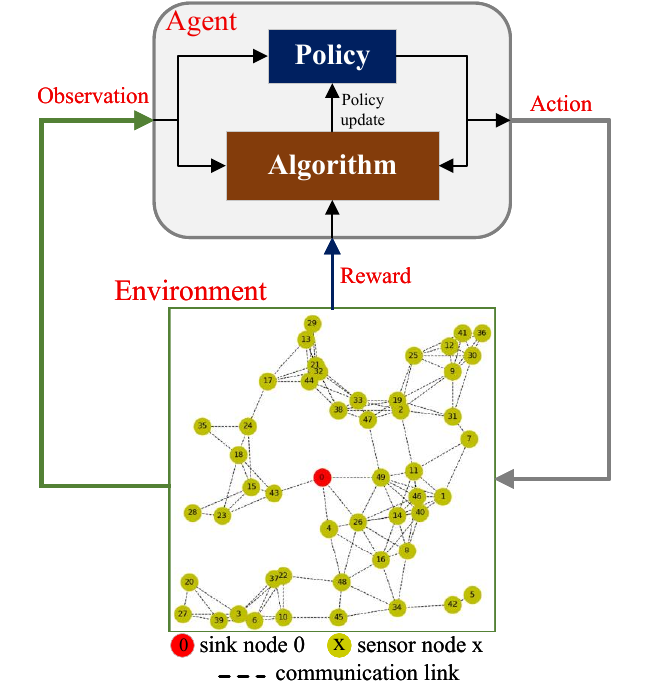}
\caption{An overall reinforcement learning model.}
\label{fig:rl_model}
\end{figure}

This section presents a Q-learning framework for time-critical data aggregation scheduling in IoT networks, as illustrated in Fig.~\ref{fig:rl_model}. By leveraging reinforcement learning, the framework dynamically assigns interference-free time slots to sensor nodes, minimizing aggregation delay $D$. The scheduling problem is modeled as an MDP ~\cite{sutton1998reinforcement}, with states, actions, rewards, and transitions designed to optimize batch transmissions. Through iterative learning, the framework converges to a near-optimal policy that reduces the number of time slots while ensuring collision-free data aggregation, particularly effective in asymmetric topologies like corner sink configurations.

Fig.~\ref{fig:rl_model} depicts the reinforcement learning model, showing the interaction between the scheduler (agent) and the IoT network (environment). The MDP components include states (representing aggregated node sets), actions (selecting initial senders), rewards (promoting large, interference-free batches), and transitions (updating aggregated nodes). This model enables the agent to learn efficient schedules by balancing exploration and exploitation, achieving lower delays than the traditional heuristic.

\subsection{Markov Decision Process (MDP) Formulation}
The data aggregation scheduling is formulated as an MDP to enable the agent to learn optimal decisions through trial and error.

\subsubsection{State Representation}

We address the aggregation scheduling problem by jointly constructing the aggregation structure - typically represented as a tree rooted at the sink node - and assigning transmission time slots to the nodes as they are incorporated into the tree. The transmission time slot of a node is determined by its position in the overall transmission order. Within any given slot, multiple nodes may transmit concurrently, provided that their transmissions are collision-free.

A scheduling solution may determine this transmission order in either a bottom-up or a top-down manner \cite{nguyen2024adaptive}. This study seeks to model and learn the top-down scheduling behavior. Specifically, the process begins by identifying the node that transmits last, followed by those that transmit earlier. Formally, if the aggregation delay is denoted by $D$, the algorithm sequentially determines the transmissions occurring at time slots $D, D-1, D-2, \ldots, 1$, completing the process in $D$ steps. In other words, the algorithm identifies the set $S_D, S_{D-1},
\dots,S_1$, in that order. Given the initial state $s_0=\{v_0\}$, the network state at step $k\in [1,D]$, i.e., $s_k$ is defined as follows:

\begin{equation}
    s_k = s_{k-1}\cup S_{D-(k-1)}.
\end{equation}
In practice, however, the actual value of $D$ is unknown until the scheduling process is complete. Consequently, once the scheduling procedure has identified all transmitting orders of nodes, the resulting schedule must be reversed to obtain the actual transmission order.



\begin{figure}[tb!]
    \centering
    \includegraphics[width=\linewidth]{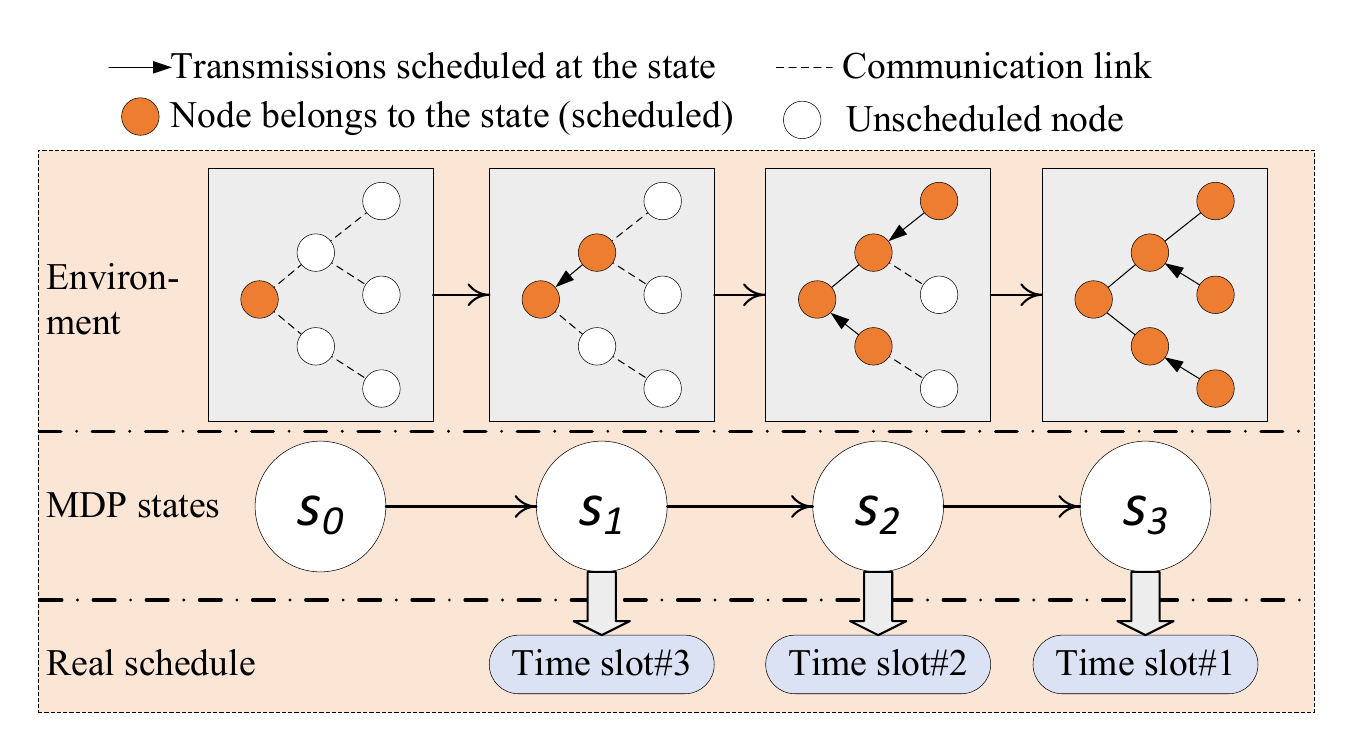}
    \caption{Example of network state transitions. State $S_0$ contains a root node. Next state includes all nodes from previous states.}
    \label{fig:mdp}
\end{figure}

Fig.~\ref{fig:mdp} illustrates an example MDP consisting of four states, corresponding to an aggregation delay of three time slots. The initial state, $s_0$, contains only the sink node. The subsequent state, $s_1$, includes the sink and the node that is determined to transmit data to the sink at time slot~3. The network then transitions from $s_1$ to $s_2$ after identifying two additional transmitters, which, in the normal schedule, transmit at time slot~2. Finally, the network transitions to $s_3$ by adding the remaining unscheduled sensor nodes to the aggregation tree. These nodes transmit their data at time slot~1. At this stage, the scheduling process is complete, and the MDP has reached its terminal state.

As presented previously, in general, the aggregation delay $D$ is not known \textit{a priori} and is not deterministic. Therefore, the scheduler must iteratively determine the network states $s_0, s_1, \ldots, s_{D}$ until the final state is reached. Once all states have been identified, the transmission order must be reversed to obtain the actual schedule: nodes in $S_{D}$ transmit at time slot~1, nodes in $S_{D-1}$ transmit at time slot~2, and so on, with nodes in $S_1$ transmitting at the final time slot~$D$. From this point onward, we refer to a node as \textit{scheduled} (or \textit{aggregated}) once it has been included in any determined state, although the actual transmission order is obtained only after the final reversal.  

For clarity, the actions that trigger state transitions are omitted in this illustrative figure. These actions will be described in the following subsection. A different selected action at each state may result in different state trajectory, and consequently, result in different aggregation delay $D$.

\subsubsection{Action Space}

\begin{figure}[tb!]
    \centering
    \includegraphics[width=.7\linewidth]{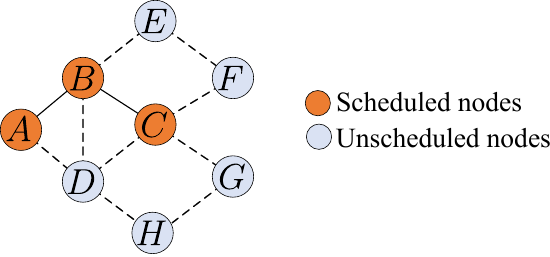}
    \caption{Example topology with scheduled and unscheduled nodes. Dashed lines are communication links.}
    \label{fig:batch_matching}
\end{figure}


At each state, given the aggregated set ($V_s$) and non-aggregated (unscheduled) set $V_{ns}$, the algorithm needs to find the set of senders in $V_{ns}$ and their corresponding receivers in ($V_s$). We define an action as the selection of an initial sender from non-aggregated nodes ($V_{ns}$) which are adjacent to at least one node in $V_s$. Subsequent senders and aggregators are chosen greedily to maximize the batch transmissions while avoiding interference. For example, in Fig.~\ref{fig:batch_matching}, assume that node~$A$ is the sink (root), and in the current state $V_s = \{A, B, C\}$. Among the non-aggregated nodes $V_{ns} = \{D, E, F, G, H\}$, the nodes adjacent to $V_s$ are $\{D, E, F, G\}$, i.e., the action space of the current state. If node~$D$ is selected as the initial sender, then all nodes $E, F, G$ are ineligible to transmit toward the aggregated nodes due to collisions at either $B$ or $C$. In contrast, if node~$E$ is selected first, then either node~$F$ or node~$G$ (but not both) can also be chosen as a sender: node~$E$ sends data to~$B$, while node~$F$ (or~$G$) sends data to~$C$.

The episode terminates when no valid actions remain ($|V_{ns}| = 0$), indicating all data has reached the sink. The initial sender and subsequent senders form a set called $T$. These senders are supposed to send data at the same time slot, and the size of $T$ more or less depends on the initial sender selection, as explained with Fig. \ref{fig:batch_matching}.

\subsubsection{Reward Function}
The reward function promotes efficient, collision-free batch transmissions by incentivizing the selection of larger sender sets. For each action, specifically the initial sender choice, the reward is defined as $r = |T|^2$, where $T$ represents the set of nodes transmitting within a single time slot. The quadratic formulation enhances the advantage of larger batches, thereby accelerating the data aggregation process. The greedy selection strategy ensures feasible, interference-free choices, balancing immediate progress with long-term delay minimization by maximizing batch sizes.

\subsubsection{Transition Dynamics}
Upon execution of an action, the state transitions by adding the batch of successfully aggregated nodes ($T$) to the aggregated set ($V_s$), with $V_{ns}$ accordingly reduced. The next state is computed deterministically as the hashed string of the updated $V_s$. The episode ends when $V_{ns}$ is empty, signifying complete data aggregation.

\subsection{Q-Learning Algorithm}

\begin{algorithm}[htbp]
\small
\SetAlgoLined
\SetKwInOut{Input}{Input}
\SetKwInOut{Output}{Output}
\Input{Graph $G = (V, E)$, number of episodes, learning rate $\alpha$, discount factor $\gamma$, initial exploration rate $\epsilon$, exploration decay rate $\epsilon_{\text{decay}}$}
\Output{Best Q-table $Q_{\text{best}}$ and minimum delay $D_{\text{best}}$}

Initialize Q-table as empty nested dictionary\;
Set best delay $D_{\text{best}} \gets \infty$, best Q-table $Q_{\text{best}} \gets \emptyset$\;
\For{each episode}{
    \tcc{Reset environment for a new episode}
    Aggregated nodes: $V_s \gets \{v_0\}$\;           
    Non-aggregated nodes: $V_{ns} \gets \{v_1, \dots, v_N\}$\;
    Current state: $s \gets \text{hash}(V_s)$\;
    delay $\gets 0$\;                             
    \While{not done}{
        \eIf{random() $<$ $\epsilon$}{
            \tcc{Exploration: Select a random sender}
            action $a \gets u \in V_{ns} \big| N(u) \cap V_s \neq \emptyset$\;
        }{
            \tcc{Exploitation: Select sender with highest Q-value}
            action $a \gets \arg\max\limits_{a'} Q(s, a')$ (break ties by lowest node ID)\;
        }
        \tcc{Execute action with greedy spread to optimize batch}
        Initialize transmitter (i.e., sender) set: $T \gets \{a\}$\;
        Filter the candidate sender set: $T_{cand} \gets \{ u \in V_{ns} \setminus \{a\} \Big| N(u) \cap V_s \neq \emptyset\}$\;
        Filter the candidate aggregator set: $C_{cand} \gets \{u \in V_s \mid N(u) \cap V_{ns} \neq \emptyset\}$\; 
        $v \gets \arg\min\limits_{u \in N(a)\cap C_{cand}} |N(u) \cap V_{ns}|$\; 
        Initialize aggregator set: $C \gets \{v\}$ 
        
        Collision\_Remover($a, v, T_{cand}, C_{cand}$)\;
        \While{$T_{cand} \neq \emptyset$}{
            \tcc{Greedy spread to select more senders and aggregators}
            $u \gets \arg\min\limits_{v \in T_{cand}} |N(v) \cap V_s|$\; 
            $r \gets \arg\min\limits_{v \in N(u)\cap C_{cand}} | N(v) \cap V_{ns}|$ \; 
            $T \gets T \cup \{u\}$\;
            $C \gets C \cup \{r\}$\;
            Collision\_Remover($u, r, T_{cand}, C_{cand}$)\;
        }
        Compute reward: $r \gets |T|^2$\;
        Update $V_s \gets V_s \cup T$, $V_{ns} \gets V_{ns} \setminus T$\;
        Next state: $s' \gets \text{hash}(V_s)$, done $\gets (|V_{ns}| = 0)$\; 
        \BlankLine
        Update $Q(s, a)$ using Eq.~\ref{eq:bellman}\;
        Current state: $s \gets s'$, delay $\gets \text{delay} - 1$\;
    }
    \BlankLine
    \tcc{Update best solution if current delay is lower}
    \If{$-\text{delay} < D_{\text{best}}$}{
        $D_{\text{best}} \gets -\text{delay}$, $Q_{\text{best}} \gets Q$\;
        Save $Q_{\text{best}}$\;
    }
    $\epsilon \gets \max(\epsilon - \epsilon_{\text{decay}}, 0)$\;
}
\Return $Q_{\text{best}}$
\caption{Q-Learning Training Phase}

\label{alg:qlearning}
\end{algorithm}

\begin{algorithm}[t]
\small
\SetAlgoLined
\SetKwInOut{Input}{Input}
\SetKwInOut{Output}{Output}
\Input{Sender node $s$, Receiver node $r$, transmitter candidates $T_{cand}$, collector candidates $C_{cand}$}
\Output{Updated $T_{cand}$ and $C_{cand}$}
\BlankLine
$T_{cand} \gets T_{cand} \setminus \{s\}$\; 
$T_{cand} \gets T_{cand} \setminus \{u \in T_{cand} \mid u \in N(r)\}$\;
$C_{cand} \gets C_{cand} \setminus \{r\}$\;
$C_{cand} \gets C_{cand} \setminus \{u \in C_{cand} \mid u \in N(s) \}$\;

\BlankLine
\Return{$T_{cand}$, $C_{cand}$}
\caption{Collision removal in the same time slot}
\label{alg:remove_collisions}
\end{algorithm}

Algorithm~\ref{alg:qlearning} outlines the Q-learning process, which iteratively refines the scheduling policy over multiple episodes. The Q-table, initialized as an empty nested dictionary, maps state hashes to action-value pairs. Key hyperparameters include the learning rate $\alpha$, discount factor $\gamma$, and exploration probability $\epsilon$ with linear decay, as specified in Section~\ref{exp_results}. The Q-table expands dynamically for unseen states, and random seeds ensure reproducibility.

At each time step, an $\epsilon$-greedy policy selects an action from nodes in $V_{ns}$ adjacent to $V_s$: with probability $\epsilon$, a random sender is chosen, otherwise the sender with the highest Q-value is selected, with ties broken by the lowest node ID. For unseen states, a random action initializes a new Q-table entry. The selected action initiates a greedy spread process to determine the sender set $T$ and aggregator set $C$, beginning with the initial sender and iteratively incorporating candidates based on the least number of neighbors in $V_{ns}$ or $V_s$. Collisions are prevented using the collision removal procedure outlined in Algorithm~\ref{alg:remove_collisions}, which removes the picked sender from further consideration and excludes neighboring nodes that could cause primary or secondary interference. This process yields the next state $s'$, reward $r = |T|^2$, and a done flag. The Q-value is updated using the Bellman equation:
\begin{equation}
Q(s, a) \leftarrow Q(s, a) + \alpha \left[ r + \gamma \max_{a'} Q(s', a') - Q(s, a) \right],
\label{eq:bellman}
\end{equation}
where $s$ is the current state, $a$ the initial action, $s'$ the next state, and $\max_{a'} Q(s', a')$ is the maximum Q-value for the next state (set to 0 if terminal). The exploration probability $\epsilon$ decays linearly to favor exploitation as episodes progress. An episode terminates when all nodes are aggregated ($|V_s| = N$). The best Q-table and minimum delay, derived from the negative of the current delay, are updated and saved whenever a new minimum delay is achieved, with post-training evaluation using $\epsilon = 0$ to compute the final schedule and reverse the schedule to positive values.

\subsection{Scheduling Process}

The proposed algorithm unifies aggregation tree construction and scheduling into a single, integrated scheduling process for static IoT networks, leveraging Q-learning during the training phase with greedy heuristics to develop an optimal scheduling policy stored in the learned Q-table. After training, this policy is applied to generate interference-free schedules, where the initial sender for each time slot is selected based on the highest Q-value for the current state, reflecting the batch transmissions learned during training. The resulting interference-free schedule implicitly forms an aggregation tree, adhering to dependency constraints, where nodes transmit only after their children. This unified approach effectively addresses the challenges of multi-hop static IoT networks, including asymmetric configurations like sink corner setups, where longer paths to the sink, increased collision risks leading to complications in latency minimization and interference management. Performance enhancements are detailed in Section~\ref{exp_results}.

\section{Experiments} \label{exp_results}
\subsection{Settings}
 
The experiments were conducted on 30 random topologies for each configuration, with the sink positioned at either the center or a corner of a $100 \times 100$ deployment area. Topologies were randomly generated using Python processed using NetworkX for connectivity analysis and visualization. Table~\ref{tab:settings} summarizes the network and Q-learning parameters, including the learning rate ($\alpha = 0.1$) for balancing exploration and exploitation, the discount factor ($\gamma = 0.9$) for prioritizing long-term rewards, and the exploration decay ($1/\text{episodes}$) to ensure convergence over 20,000 episodes per topology.

\begin{table}[htbp]
\caption{Network and Model Parameter Settings}
\label{tab:settings}
\centering
\begin{tabular}{l|c}
\toprule
\textbf{Parameter} & \textbf{Value} \\
\midrule
Network area & $100 \times 100$ $unit^2$ \\
Number of nodes ($N$) & 50, 100, 150, 200, 250, 300 \\
Sink position & Center, Corner \\
Communication range ($R$) & 20 $units$ \\
Number of topologies & 30 per configuration \\
Learning rate ($\alpha$) & 0.1 \\
Discount factor ($\gamma$) & 0.9 \\
Initial exploration probability ($\epsilon$) & 1.0 \\
Number of episodes & 20,000 per topology \\
Exploration decay ($\epsilon_{\text{decay}}$) & $1/\text{(Number of episodes)}$ \\
\bottomrule
\end{tabular}
\end{table}

\subsection{Experimental Results}

The proposed Q-learning framework was evaluated against the break-and-join heuristic~\cite{nguyen2020break}, with the aggregation delay $D$ defined as the number of time slots required for all data to reach the sink. The results, averaged over 30 random topologies per configuration, are presented in Table~\ref{tab:results_center} for the central sink and Table~\ref{tab:results_corner} for the sink corner cases. For the sink center, Q-learning achieves 3.00-10.13\% lower delay than the heuristic, while for the sink corner, improvements range from 0.85-10.87\%, reflecting superior adaptation to asymmetric topologies. 

\begin{figure}[htbp]
\centering
\begin{tabular}{c@{\hspace{1mm}}c}
\includegraphics[width=0.515\columnwidth]{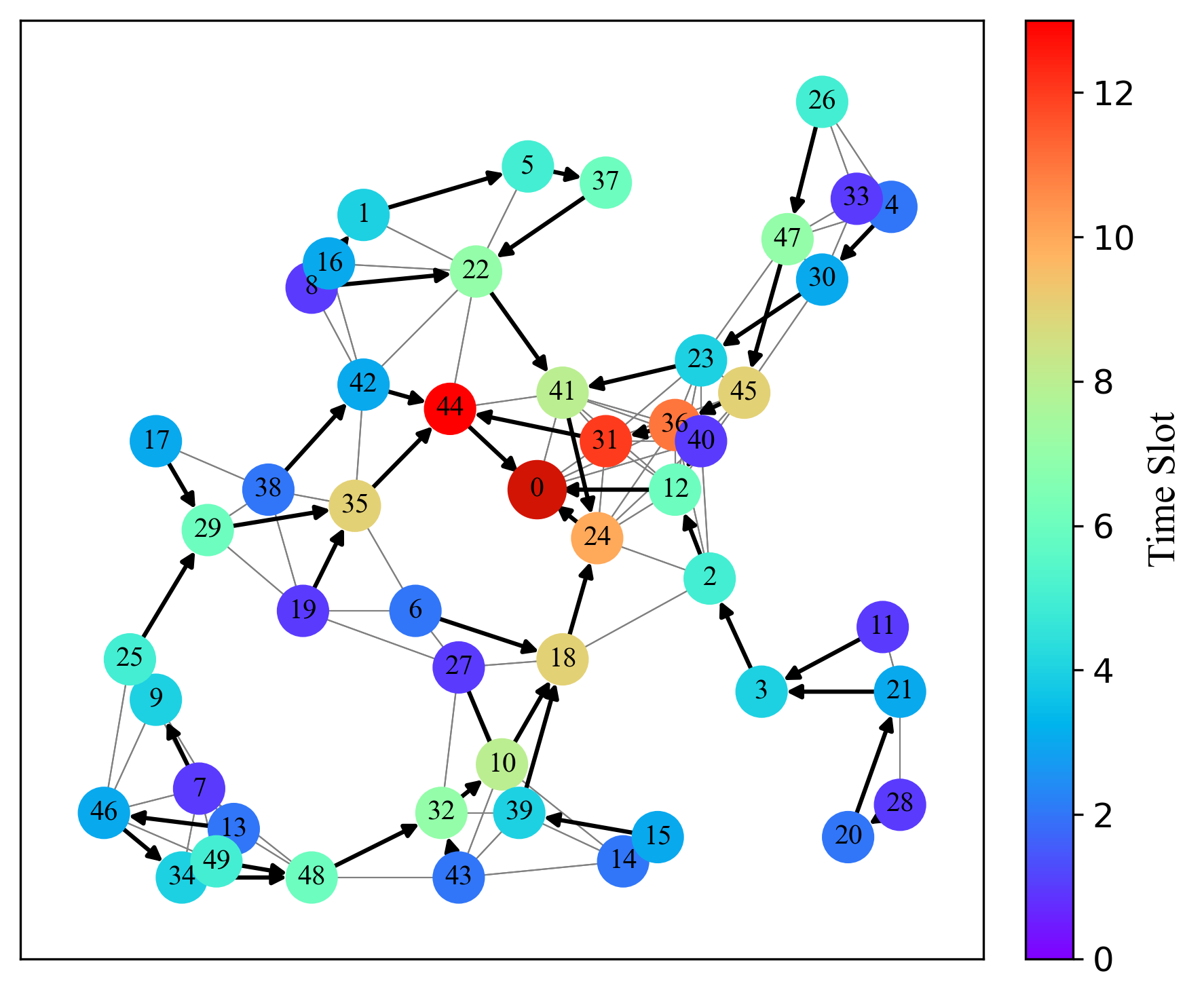} & \includegraphics[width=0.425\columnwidth]{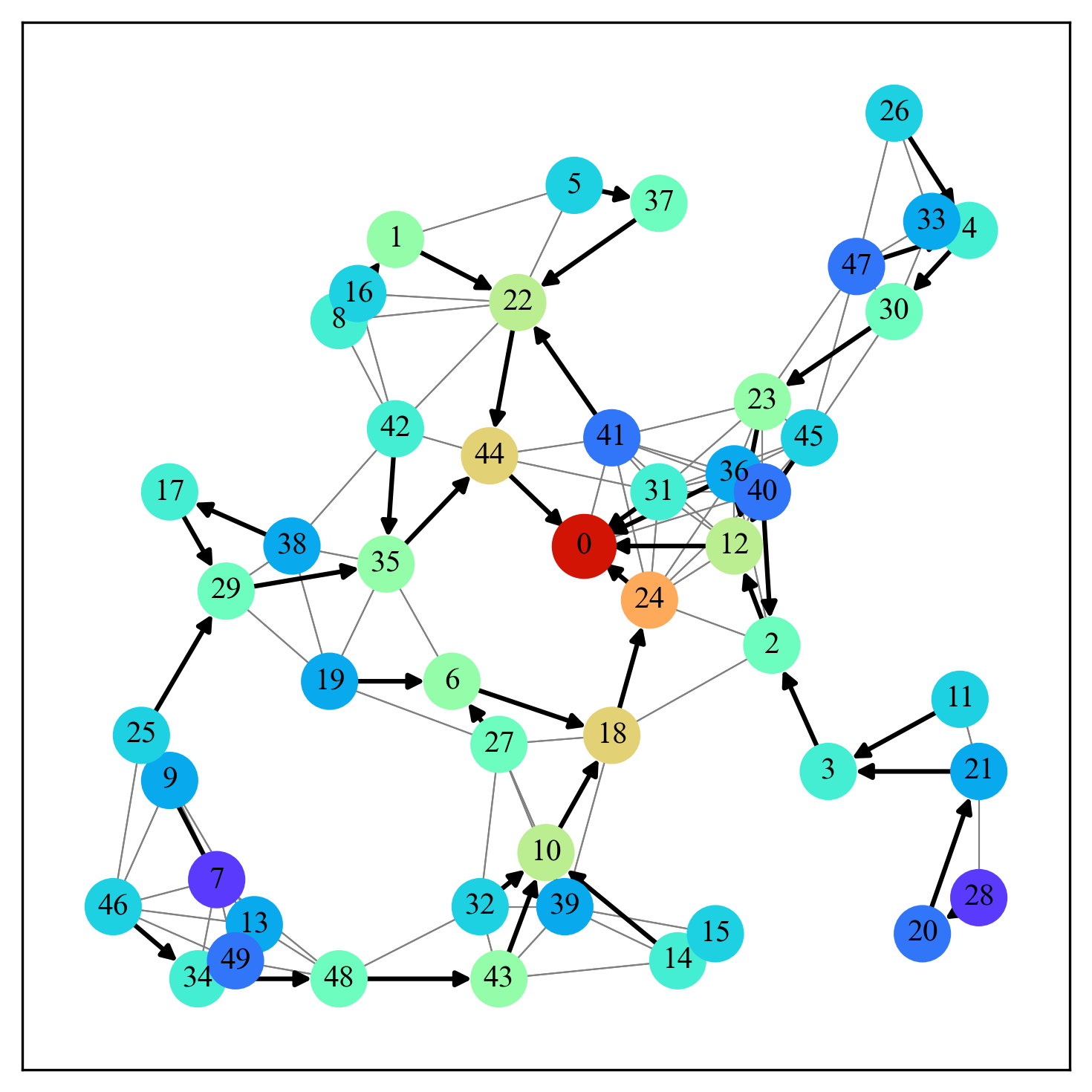} \\
(a) Heuristic & (b) Q-learning \\
\end{tabular}
\caption{Data aggregation schedule for a 50-Node Network (Sink center).}
\label{fig:aggre_50}
\end{figure}

\begin{figure}[htbp]
\centering
\begin{tabular}{c@{\hspace{1mm}}c}
\includegraphics[width=0.515\columnwidth]{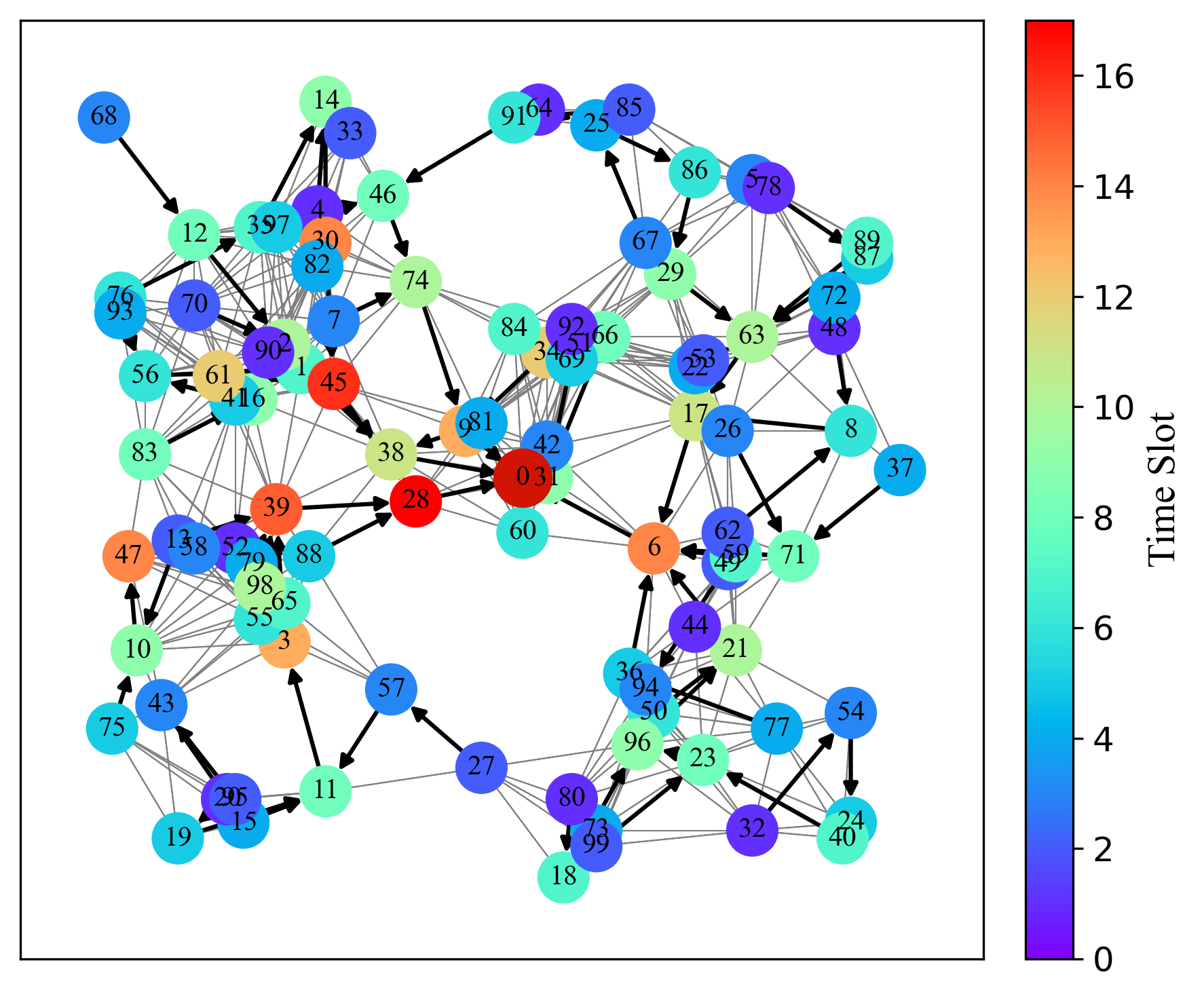} & \includegraphics[width=0.425\columnwidth]{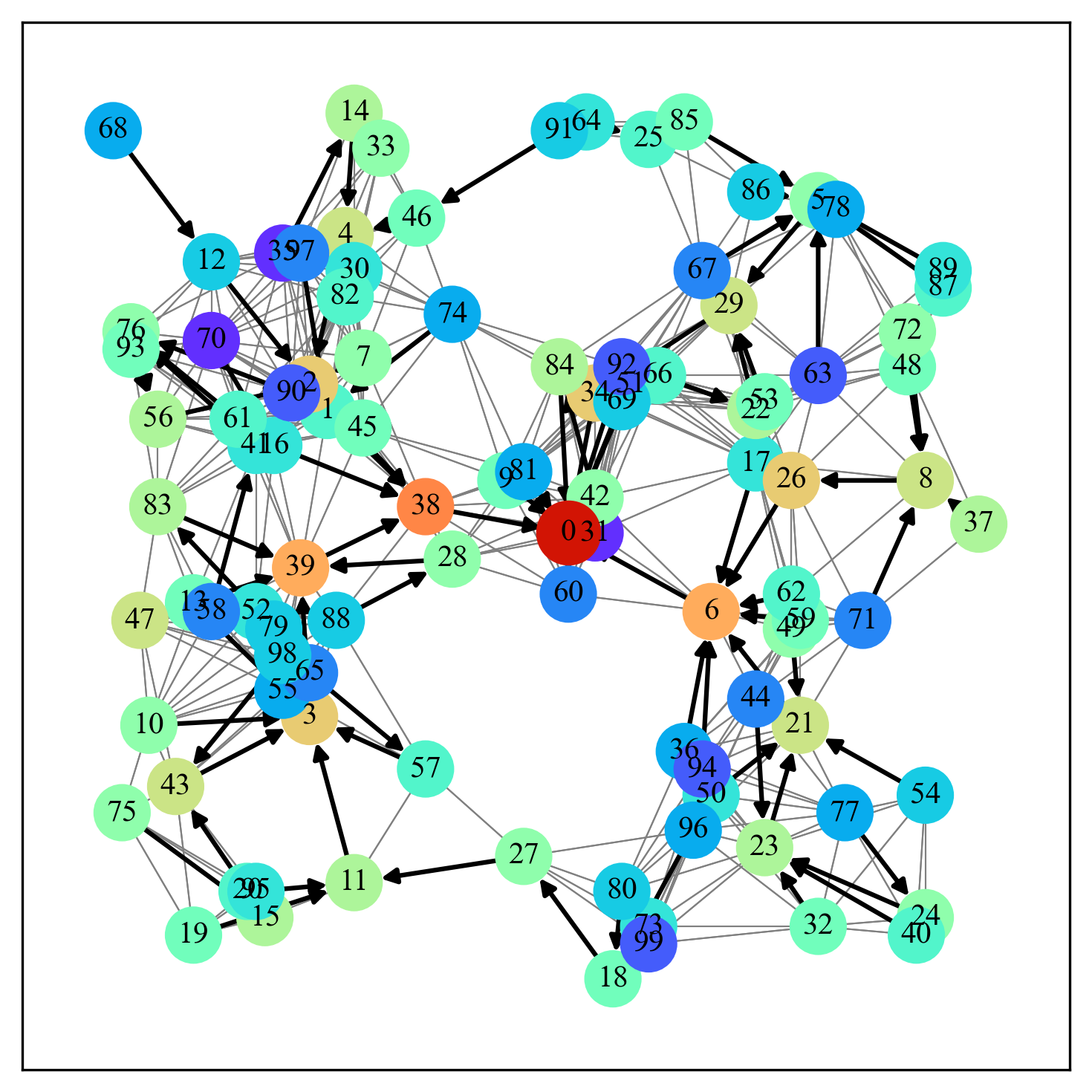} \\
(a) Heuristic & (b) Q-learning \\
\end{tabular}
\caption{Data aggregation schedule for a 100-Node Network (Sink center).}
\label{fig:aggre_100}
\end{figure}

Fig.~\ref{fig:aggre_50} illustrates the data aggregation schedules for a 50-node network (sink center), with (a) Heuristic requiring 13 time slots and (b) Q-learning requiring only 10 time slots. Nodes are colored by time slot (sink in red, others in rainbow colormap), with solid edges (with arrows) indicating the aggregation tree and dashed edges showing neighbor connections within $R=20$. Similarly, Fig.~\ref{fig:aggre_100} shows schedules for a 100-node network, where Q-learning reduces the delay to 14 time slots compared to 17 for the heuristic, demonstrating scalability in denser networks. Fig.~\ref{fig: percentage} plots the percentage delay improvement of Q-learning over the heuristic across network sizes (50-300 nodes), with (a) sink center and (b) sink corner, highlighting consistent performance gains, especially in asymmetric topologies. In sparse networks, the heuristic performs adequately due to fewer collisions, but in dense networks, its predefined tree construction wastes time slots by failing to optimize batch transmissions. Q-learning, leveraging its exploration phase, dynamically adjusts to dense topologies, achieving near-optimal collision-free schedules.

\begin{table}[htbp]
\caption{Average Aggregation Delay (Sink center)}
\label{tab:results_center}
\centering
\begin{tabular}{c|c|c}
\toprule
\textbf{Nodes} & \textbf{Heuristic (time slot)} & \textbf{Q-Learning (time slot)} \\
\midrule
50  & 10.87 & \textbf{10.2} \\
100 & 15.10 & \textbf{13.57} \\
150 & 18.60 & \textbf{17.43} \\
200 & 23.13 & \textbf{21.40} \\
250 & 26.43 & \textbf{25.3} \\
300 & 29.97 & \textbf{29.07} \\
\bottomrule
\end{tabular}
\end{table}

\begin{table}[htbp]
\caption{Average Aggregation Delay (Sink corner)}
\label{tab:results_corner}
\centering
\begin{tabular}{c|c|c}
\toprule
\textbf{Nodes} & \textbf{Heuristic (time slot)} & \textbf{Q-Learning (time slot)} \\
\midrule
50  & 15.33 & \textbf{15.20} \\
100 & 18.90 & \textbf{17.03} \\
150 & 19.20 & \textbf{17.57} \\
200 & 27.23 & \textbf{24.27} \\
250 & 26.67 & \textbf{25.37} \\
300 & 34.10 & \textbf{31.83} \\
\bottomrule
\end{tabular}
\end{table}




\subsection{Discussion}

\begin{figure}[!htb]
\centering
\begin{tabular}{c@{\hspace{1mm}}c}
    \includegraphics[width=.45\columnwidth]{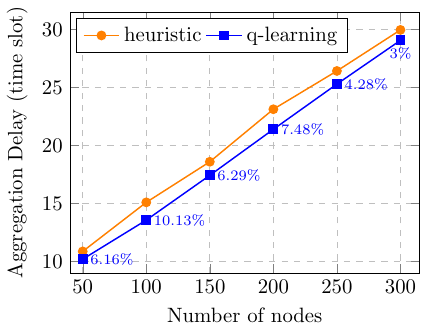} &
    \includegraphics[width=.45\columnwidth]{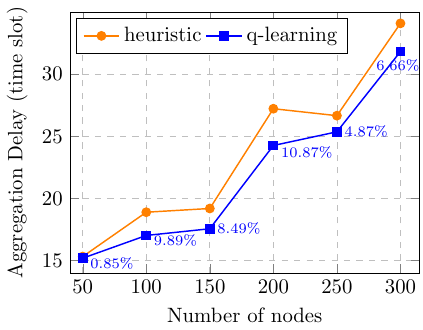} \\
    (a) Sink center & (b) Sink corner \\
    \end{tabular}
    \caption{Delay improvement of Q-learning approach in different network sizes.}
    \label{fig: percentage}
\end{figure}

{\sloppy 
The Q-learning framework consistently outperforms the break-and-join heuristic~\cite{nguyen2020break}, achieving delay reductions of up to 10.13\% in sink center configurations and 10.87\% in sink corner configurations, as shown in Fig.~\ref{fig: percentage}. The superior performance in sink corner scenarios highlights Q-learning’s ability to adapt to asymmetric topologies, where longer paths to the sink challenge heuristic methods. In sparse networks, the heuristic performs adequately due to fewer collisions, but in dense networks, its static tree construction leads to suboptimal scheduling. In contrast, Q-learning leverages its exploration phase to dynamically select large, collision-free batches, achieving near-optimal delays after approximately 20,000 episodes. However, the framework’s reliance on static topologies limits its applicability to dynamic IoT scenarios. Future work will address this by incorporating node mobility, integrating energy constraints into the reward function, and exploring deep reinforcement learning for larger-scale networks, enhancing robustness for real-world, time-sensitive IoT applications.
}

\section{Conclusion} \label{conclusion}

This paper presents a Q-learning-based framework for time-critical data aggregation scheduling in IoT networks, effectively integrating aggregation tree construction and scheduling into a unified MDP. By employing hashed state representations and a reward function that promotes large, interference-free batch transmissions, the approach achieves up to 10.87\% lower latency compared to the state-of-the-art break-and-join heuristic. Simulations on 30 random topologies with 50 to 300 nodes, for both center and corner sink positions, demonstrate the framework's robustness in static IoT environments. The greedy-augmented action selection mechanism ensures efficient, collision-free schedules, particularly in asymmetric topologies.

Despite its effectiveness, the current framework assumes static network topologies, limiting its applicability to dynamic IoT scenarios with node mobility or failures. Future work will focus on extending the approach to handle dynamic topologies by incorporating adaptive state representations and exploring deep reinforcement learning for larger-scale networks. Additionally, integrating energy constraints into the reward function will enhance the framework's suitability for resource-constrained IoT devices. These advancements aim to further improve latency and efficiency in real-world, time-sensitive IoT applications.


\printbibliography

\end{document}